\documentclass[twocolumn,showpacs,preprintnumbers,amsmath,amssymb,superscriptaddress]{revtex4}

\usepackage{graphicx}
\usepackage{graphicx,epsf} 
\usepackage{dcolumn}
\usepackage{bm}
\usepackage{latexsym}

\newcommand{\xsize}{\epsfxsize=8.5cm}

\begin{document}

\title{Graph Partitioning Induced Phase Transitions}

\author{Gerald~Paul}
\affiliation{Center for Polymer Studies and Dept.\ of Physics, Boston
  University, Boston, MA 02215, USA} 
\email{gerryp@bu.edu}

\author{Reuven~Cohen}
\affiliation{Center for Polymer Studies and Dept.\ of Physics, Boston
  University, Boston, MA 02215, USA} 
\affiliation{New England Complex Systems Institute, Cambridge, MA 02138, USA}

\author{Sameet Sreenivasan}
\affiliation{Center for Polymer Studies and Dept.\ of Physics, Boston
  University, Boston, MA 02215, USA} 
\affiliation{Dept. \ of Physics, University of Notre Dame, Notre Dame, IN 46556, USA}

\author{Shlomo Havlin}
\affiliation{Center for Polymer Studies and Dept.\ of Physics, Boston
  University, Boston, MA 02215, USA} 
\affiliation{ Minerva Center and Department of Physics, Bar Ilan
University, Ramat Gan 52900, Israel}

\author{H.~Eugene Stanley}
\affiliation{Center for Polymer Studies and Dept.\ of Physics, Boston
  University, Boston, MA 02215, USA} 
 
\begin{abstract}

  We study the percolation properties of graph
  partitioning on random regular graphs with N vertices of degree $k$.
  Optimal graph partitioning is directly related to optimal attack and
  immunization of complex networks.  We find that for any partitioning
  process (even if non-optimal) that partitions the graph into equal sized
  connected components (clusters), the system undergoes a percolation phase
  transition at $f=f_c=1-2/k$ where $f$ is the fraction of edges removed to
  partition the graph.  For optimal partitioning, at the percolation
  threshold, we find $S \sim N^{0.4}$ where $S$ is the size of the clusters
  and $\ell\sim N^{0.25}$ where $\ell$ is their diameter.  Additionally, we
  find that $S$ undergoes multiple non-percolation transitions for $f<f_c$.

\end{abstract}  

\maketitle


The graph partitioning problem deals with assigning vertices in a graph to
different partitions such that no partition is greater than a given size.
The optimal solution is one which minimizes the fraction of edges $f$
that must be removed such that there are no edges between partitions
\cite{Kernighan}.

Graph partitioning is of interest not only because of the large amount of
previous research done but also because optimal partitioning is equivalent to
optimal attack/immunization of a complex network.  That is, the percolation
threshold $f_c$, at which global connectivity is lost, will be lower than
that for any other type of attack/immunization and the measure of fragmentation $F$
\cite{Borgatti} for all values of $f$ will be higher than for any other type
of attack/immunization \cite{Holme}.

Graph partitioning is a much studied subject with a long history of work by
mathematicians and computer scientists. The problem of determining the
optimal solution is NP complete.  Mathematicians have pursued finding
rigorous bounds for the minimum number of edges needed to partition (usually
random regular) graphs into two equal sized partitions
\cite{Buser,Bollabas,Alon,Bezrukov}.  Computer scientists have pursued
developing efficient algorithms which heuristically find good approximations
to the optimal solution \cite{Kernighan,Boettcher,Karypis1,Metis}.

Here we study graph partitioning from the standpoint of statistical
physics.  To make contact with percolation theory \cite{Stauffer,
  BundeHavlin}, we identify the number of edges removed as the control
variable and study the inverse problem: given that we are allowed to remove a
fraction $f$ of the edges from the graph, how can we partition the graph to
minimize the size of the largest partition.  We denote as $S$ the size of the 
largest connected component (cluster) which results from the partitioning.  Then, 
$S$ plays the role of
order parameter and we are interested in the behavior of S as a function of
$f$.  We ask if there is a critical value $f_c$ such that for $f <f_c$, $S
\sim N$ while for $f > f_c$, $S$ scales slower than $O(N)$.  That is, does the
graph undergo a percolation phase transition? If so, what is the percolation
threshold $f_c$ and what are the critical exponents associated with the phase
transition.

We study random $k-regular$ graphs, random graphs the vertices of which all
have the same degree, $k$. We study these graphs because of their intrinsic
interest and because these graphs are examples of expander graphs which are
extremely robust to node or edge removal \cite{expander1, expander2}.  They are
therefore a good testbed for optimal graph partitioning.

We find that, in fact, a percolation transition does exist and we analytically
determine $f_c$.  We also estimate critical exponents associated with the
transition.  In addition however, we find that for $f<f_c$ the graph
undergoes a large number of first order transitions related to the partitioning
process.


{\it Percolation Threshold.} The percolation threshold can be determined
analytically as follows.  In Refs.  \cite{Cohen2000,Cohen2002} it was argued
that for a random graph having a degree distribution $P(k)$ to have a
spanning cluster, a vertex $j$ which is reached by following a link (from vertex 
$i$ on) the giant cluster must have at least one other link, on average to
allow the cluster to exist.  Or, given that vertex $i$ is connected to $j$, the
average degree of vertex $j$ must be at least 2:
\begin{equation}
<k_i|i \leftrightarrow j>=2.
\label{k2}
\end{equation}

We will show below that, for large $N$ at the percolation threshold, all
partitions are essentially the same size and that each partition consists of
one cluster \cite{note0}.  Then, to achieve Eq.~(\ref{k2}) the average degree
in each cluster must be 2 and $p_c$ the fraction of edges which must be
present is
\begin{equation}
p_c \equiv 1-f_c=\frac{2}{k}.
\label{pc}
\end{equation}
This is to be compared to the random site or bond percolation threshold
$p_c=1/(k-1)$ \cite{Cohen2000}.

We can gain insight into the structure of the spanning clusters by noting
that for tree graphs with $n$ vertices
\begin{equation}
<k>=\frac{2(n-1)}{n}
\end{equation}
which approaches 2 as $n \rightarrow \infty$.  For finite graphs, however, to
satisfy $<k>=2$, there must be on average one loop in each graph. Thus, at
the percolation threshold, the clusters contain on average one loop.  Our
problem can be restated as: how do we partition a graph into the largest
number of equal sized partitions each composed of one cluster with on average
one loop per cluster.  The larger the number of partitions (and thus the
smaller the partition size), the closer the solution is to the optimal one.
Different types of partitioning that maintain one cluster per partition will
result in the same critical point but the scaling of the cluster size at the
critical point may depend on the optimality of the partitioning.


{\it Optimal Partitioning.} We use the METIS graph partitioning program
\cite{Metis} which provides close to optimal graph partitioning.  For the
same random graph we run the program many times over the range of partition
sizes in which we are interested.  After each partitioning we identify the
clusters in the graph, determine the size of the largest cluster and note the
number of edges needed to be removed for the partitioning.  For each value of
the number of edges, we maintain the minimum value of the size of largest
cluster in the partitioning.

\begin{figure}[h]

\centerline{
\xsize
\epsfclipon
\epsfbox{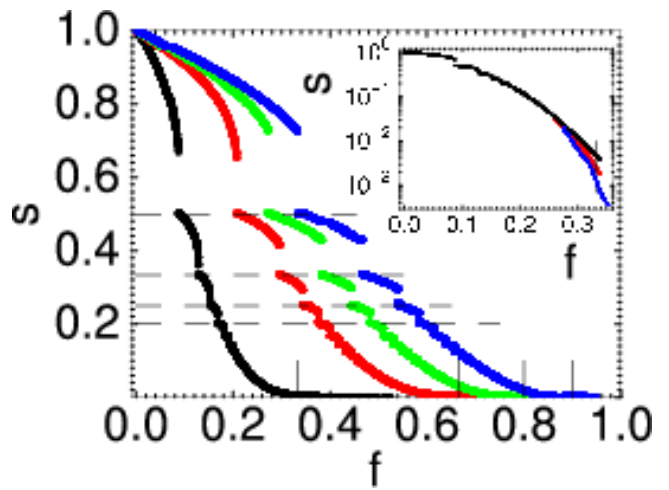}
}
\caption{Normalized largest cluster size, $s \equiv S/N$ versus fraction of edges
  removed, $f$, for random regular graphs with number of vertices $N=10^4$
  of degree (from left to right) $k=3, 6, 10$, and $20$.  The vertical lines
  at the x-axis mark the predicted values of $f_c=1-2/k$ from left to right
  for $k=3, 6, 10$, and $20$.  The dashed horizontal lines at $s=1/2, 1/3,
  1/4$, and $1/5$ are the values of $s$ for which the first few
  non-percolation transitions take place.  
Inset: For (from top to bottom on right) $N=10^4, 3 \times 10^4 $, and
  $10^5$ and $k=3$, $s$ versus $f$.  Data collapse until $f$ is in the
  vicinity of $f_c=1/3$ (indicated by vertical line).}
\label{figOptPercOverview}
\end{figure}

Figure \ref{figOptPercOverview} illustrates the behavior of $s \equiv S/N $
versus $f$ for various values of $k$ \cite{Pinf}.  In what follows we will analyze the
case $k=3$ in depth; similar results are obtained for other values of $k$.

\begin{figure}[h]

\centerline{
\xsize
\epsfclipon
\epsfbox{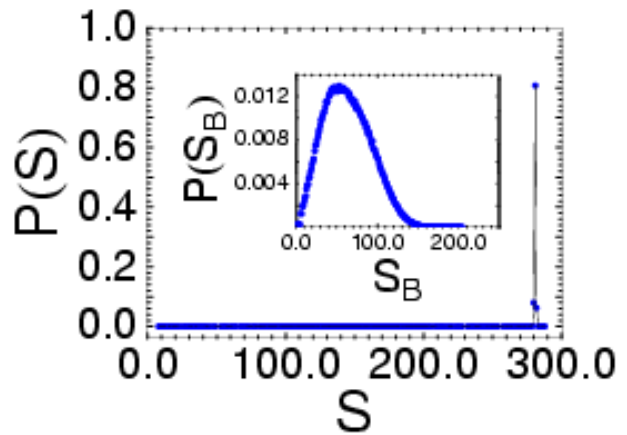}
}
\caption{For $N=10^6$ and $k=3$ at criticality, $P(S)$, the distribution of
  cluster sizes, $S$.  Inset is plot of $P(S_B)$, the
  distribution of blob sizes, $S_B$, for $N=10^6$ and $k=3$.}
\label{figOptPercSizeDists}
\end{figure}

In Fig.~\ref{figOptPercSizeDists}, for $N=10^6$, we plot $P(S)$ the
distribution of cluster sizes, $S$, versus $S$ at the threshold predicted by
Eq.~(\ref{pc}) $f_c=1/3$.  As expected, the distribution is very
strongly peaked -- almost all clusters are the same size.  In the inset in 
Fig.~\ref{figOptPercOverview} for $k=3$ and various values of $N$ we plot $s$
versus $f$.  Below $f_c$ the plots collapse indicating that here $S \sim N$.
In the vicinity of and above $f_c$ the plots no longer collapse, a
manifestation of $S$ scaling more slowly than $N$.

\begin{figure}[h]

\centerline{
\xsize
\epsfclipon
\epsfbox{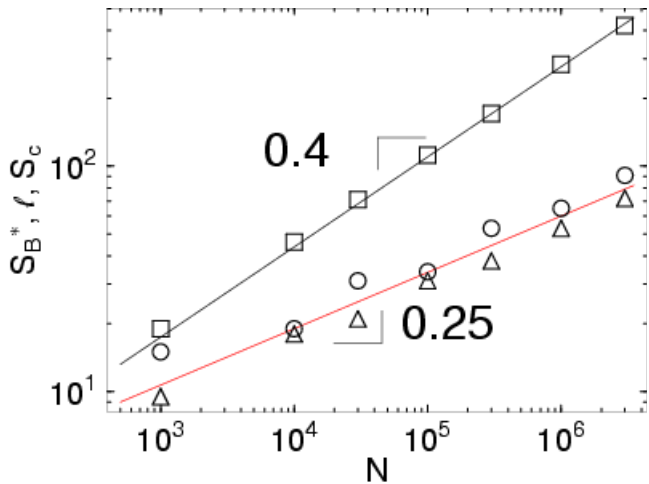}
}
\caption{Largest cluster size at criticality, $S_c$ (squares), chemical size of largest
  cluster, $\ell$ (circles), and most probable blob size $S_B^*$ (triangles),
  versus number of vertices $N$ in graph.}
\label{figOptPercScaling}
\end{figure}

\begin{figure}[h]

\centerline{
\xsize
\epsfclipon
\epsfbox{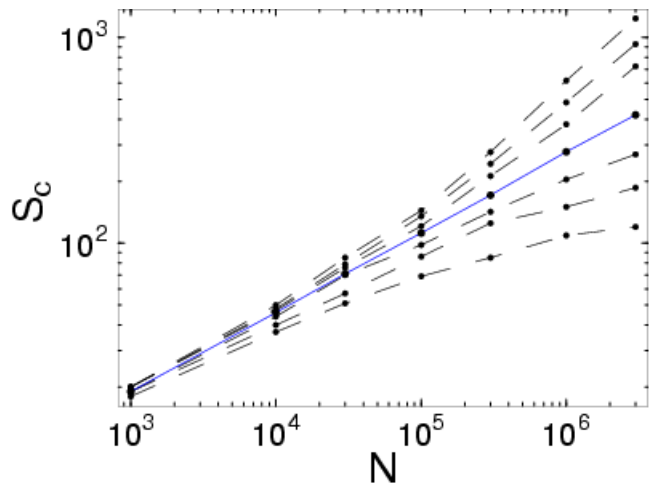}
}
\caption{Largest cluster size for (from top to bottom) values of $f=0.331,
  0.332, 0.333, 1/3$ (solid line), $0.335, 0.337$, and $0.34$, versus number of
  vertices $N$ in graph.  The straightest plot is for $f=1/3$ the predicted
  value of $f_c$}
\label{figOptPercPc}
\end{figure}

In Fig.~\ref{figOptPercScaling} we plot $S_c$ the value of S at the
percolation threshold versus $N$.  The slope of the plot is consistent with
\begin{equation}
S_c \sim N^x,
\label{SN}
\end{equation}
where $x \approx 0.4 $.  In Fig.~\ref{figOptPercPc}, we plot $S_c$ versus $N$
for various values of $f$ and see that the straightest plot is for $f_c=1/3$,
the predicted critical threshold.

\begin{figure}[h]

\centerline{
\xsize
\epsfclipon
\epsfbox{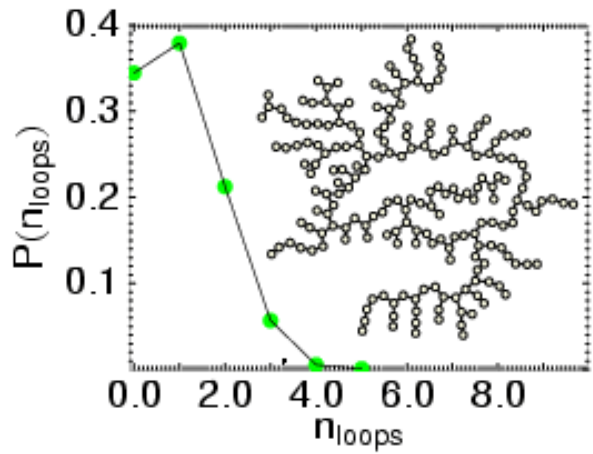}
}
\caption{At criticality for $N=10^5$ and $k=3$, $P(n_{loop})$ distribution of
  number of loops per cluster, $n_{loop}$.  Inset: For $N=10^5$ and $k=3$,
  typical cluster at criticality illustrating that typical clusters at
  criticality contain 1 loop decorated by trees.}
\label{figOptPercLoopsPerCluster}
\end{figure}

In Fig.~\ref{figOptPercScaling} we also plot $\ell$ the chemical size
(diameter) of the critical clusters versus $N$.  The slope of the plot is
consistent with
\begin{equation}
\ell \sim N^{z}
\label{elN}
\end{equation}
where $z \approx 0.25$. From Eqs.~(\ref{SN}) and (\ref{elN}) we obtain
\begin{equation}
S_c \sim \ell^{d_l}
\label{Sel}
\end{equation}
where $d_l \equiv x/z \approx 1.6$. The exponent $d_l$ is a measure of the
compactness of the clusters: clusters with $d_l=1$ are essentially chains;
higher values of $d_l$ correspond to more dense structures.  For random
percolation, $d_{\ell}=2$ \cite{BundeHavlin,HavlinNosal}.  The inset in 
Figure \ref{figOptPercLoopsPerCluster} is a representative
critical cluster obtained from partitioning.  Note the single loop required
by Eq.~(\ref{k2}) and its "stringy" structure, the manifestation of $d_l
\approx 1.6$. In Fig.~\ref{figOptPercLoopsPerCluster} we plot the
distribution of the number of loops per cluster, $P(n_{loop})$ and note that
it is fairly narrow with the most probable value being 1.  Thus, not only is
the average number of loops per cluster 1 but the most probable number is
also 1.

The exponent $\tilde \nu$ is defined by \cite{BundeHavlin}
\begin{equation}
r \sim \ell^{\tilde\nu}
\label{rlnu}.
\end{equation}
where $r$ is Euclidean distance.  At the percolation threshold, $\tilde \nu$
is expected to be $1/2$, the same value as for a random walk (or for a
network embedded in a very high dimensional lattice, such that spatial
constraints are irrelevant) \cite{BundeHavlin}.

Using Eq.~(\ref{Sel}) with $d_l=1.6$ and Eq.~(\ref{rlnu}) with $\tilde\nu=1/2$, we can
determine the fractal dimension of the percolation clusters at criticality
defined by
\begin{equation}
S_c \sim r^{d_f}
\label{srdf}
\end{equation}
to be 
\begin{equation}
d_f = \frac{ d_{\ell} }{\tilde \nu} \approx 3.2.
\end{equation}
Assuming that our problem of optimal partitioning on random regular graphs
has an analog on lattices in Euclidean space of dimension $d$, in which
\begin{equation}
N \sim r^d
\label{Nrd}
\end{equation}
where $r$ is the length of a side of the lattice, we can determine the
upper critical dimension, $d_c$ for that analog.  The upper critical
dimension is defined such that for $d \ge d_c$, all critical exponents are
unchanged.  Since random graphs can be considered to be embedded in an
infinite dimensional space, the critical exponents for our problem should be
the same as those at the critical dimension for the Euclidean analog.  Using
Eqs.~(\ref{srdf}) and (\ref{Nrd}), we find $S_c \sim N^{d_f/d_c} \sim N^{0.4}$ 
and thus $d_c = 8 $
which interestingly is the critical dimension for lattice animals and
branched polymers \cite{latticeAnimal1,latticeAnimal2}.

We can learn more about the fractal structure of the spanning cluster at
$f_c$ by analyzing the 2-connected components (blobs) \cite{HerrmannStanley}
within the spanning clusters.  This is equivalent to analyzing the loops
within the spanning clusters because the typical cluster contains 1 loop
which is the 2-connected component in the cluster.  In
Fig.~\ref{figOptPercScaling} we plot the most probable blob size (equivalent
to the length of loops), $S_B^*$, versus $N$.  The scaling is consistent
with $S_B^* \sim N^{0.25}$ similar to the scaling of the chemical length
of the whole cluster.  From this we infer that the chemical size of the
cluster is driven by the size of the loops.

{\it Non-optimal partitioning.} We find that for partitioning in which we
ensure that each partition consists of one cluster but no attempt is made to
minimize the number of edges between partitions, as predicted above, $f_c$ in
this case is also $1-2/k$ but at criticality $S \sim N^{1/2}$.  That is, the
clusters at criticality are larger than those at criticality for optimal
partitioning.  The argument that the exponent is exactly $1/2$ is as follows:
We ask how large a cluster must be to have on average one loop.  Consider a
cluster of size $S$. The total number of edges associated with vertices in
the cluster is $k S$.  Connectivity among vertices in the cluster is provided
by $2(S-1)$ of the edges and others (also of order $S$) are either removed
(connected to other partitions) or connected back to the cluster forming a
loop. Because the graph is random and we partition randomly (subject to the
constraint that the partitions consist of one cluster each), the probability
that one of these edges is connected back to the cluster is
\begin{equation}
P_{loop} \sim S \frac{S}{N}.
\label{ploop}
\end{equation}
Setting $P_{loop}=1$ we find $S  \sim N^{1/2}$. 
 
{\it Random partitioning.}  Random partitioning is achieved by assigning vertices 
to partitions randomly and is equivalent to random site percolation \cite{randomsite}, for which the 
well known result $f_c=1-1/(k-1)$ holds \cite{Cohen2000,Cohen2002}.  In contrast to the optimal 
and the non-optimal partitioning considered above, 
for random partitioning, partitions contain clusters of all sizes (including very small ones). 
Eq.~(\ref{pc}) holds for the spanning cluster in each partition but does 
not hold for all clusters and $f_c$ is therefore significantly larger.

{\it Non-percolation transitions.} In Fig.~\ref{figOptPercOverview}, we see
that the order parameter is discontinuous at values of $s={1/2, 1/3,
  \ldots}$, qualifying these points as first order phase transitions.
However, these discontinuities, which occur where the number of partitions
changes are not percolation transitions -- the scaling of $s$ with $N$ does
not change.  The behavior at these transitions (and the general shape of the
segments of the plots) can be understood as follows: Consider the region of
the plot corresponding to two partitions ($1/2 < s < 1$) and assume we reduce
the size of the larger partition (increasing the size of the smaller
partition) by moving selected vertices one-by-one from the larger partition
to the smaller partition \cite{note1}.  Initially, the number of edges needed
to be removed when we move a vertex is $k$ -- all edges adjacent to the moved
vertex must be removed.  As the size of the smaller partition increases, we
can select a vertex requiring fewer of its edges to be removed because some
of its edges already have ends in the smaller partition.  At some point, the
number of edges to the smaller partition of a vertex to be moved is equal to
the number of the vertex's edges to the larger partition -- thus, there is
zero cost to the move \cite{note2}.  This continues to be the case until the
partitions are of equal size, resulting in the discontinuity.


{\it Discussion.} If a graph with an arbitrary degree distribution
$P(k)$ can be partitioned such that there is one cluster per partition, then
our result for $f_c$ should be generalized to
\begin{equation}
f_c= 1-\frac{2}{<k>}
\end{equation}
where $<k>$ is the average degree per vertex. Areas for future work include
determining whether this is the case for partitioning on such other types of
graphs as Erdos-Renyi and scale-free graphs.  Also of interest will be
determining if there exists a Euclidean analog to our graph partitioning
problem.

We thank ONR, the Israel Science Foundation and the Dysonet Project for
support.




\begin{thebibliography}{99}
\bibitem{note0} This is a reasonable assumption because an optimal algorithm
  will not needlessly remove edges in a partition which resulting in multiple
  clusters.  If it did, the solution would not be optimal -- edges would have
  been wasted dividing the partition into multiple clusters. On the other
  hand, one cannot, a priori, rule out optimal partitioning resulting in
  multiple clusters per partition.
\bibitem{Pinf} The quantity $s$ is equivalent to the order parameter in percolation theory $P_\infty$, 
the fraction of sites in the incipient infinite cluster.
\bibitem{randomsite} Each partition is equivalent to a percolation system
  after vertices have been randomly removed from the system and is thus
  equivalent to random site percolaton.
\bibitem{note1} This is not necessarily how optimal partitioning would occur
  but it is illustrative of the point we are making.
\bibitem{note2} Here we consider graphs with even $k$.  For odd $k$, consider
  moving 2 vertices at a time.
\end{thebibliography}
\end{document}